\documentclass{ws-ijmpc}
\usepackage[super]{cite}
\usepackage{amssymb}
\usepackage{amsmath,amssymb,amsfonts}
\usepackage[table]{xcolor}
\usepackage{url}
\begin{document}
\markboth{L.K. Eraso-Hernandez, et. al.}
{Evolution of transport under cumulative damage in metro systems}

\title{Evolution of transport under cumulative damage in metro systems}
\author{L.K. Eraso-Hernandez${}^{1}$, A.P. Riascos${}^{2}$, T.M. Michelitsch${}^{3}$ and J. Wang-Michelitsch${}^4$}

\address{${}^1$Instituto de F\'isica, Universidad Nacional Aut\'onoma de M\'exico, Apartado Postal 20-364, 01000 Ciudad de M\'exico, M\'exico\\
	${}^2$ Departamento de Física, Universidad Nacional de Colombia, Bogotá, Colombia\\
	${}^3$Sorbonne Universit\'e, Institut Jean le Rond d'Alembert, CNRS UMR 7190,4 place Jussieu, 75252 Paris cedex 05, France\\
	${}^4$Independent researcher, Paris, France}

\maketitle
\begin{abstract}
	One dominant aspect of cities is transport and massive passenger mobilization which remains a challenge with the increasing demand on the public as cities grow. In addition, public transport infrastructure suffers from traffic congestion and deterioration, reducing its efficiency. In this paper, we study the capacity of transport in 33 worldwide metro systems under the accumulation of damage. We explore the gradual reduction of functionality in these systems associated with damage that occurs stochastically. The global transport of each network is modeled as the diffusive movement of Markovian random walkers on networks considering the capacity of transport of each link, where these links are susceptible to damage. Monte Carlo simulations of this process in metro networks show the evolution of the functionality of the system under damage considering all the complexity in the transportation structure. This information allows us to compare and classify the effect of damage in metro systems. Our findings provide a general framework for the characterization of the capacity to maintain the transport under failure in different systems described by networks.
\end{abstract}


\maketitle
\section{Introduction}
Metro systems represent an important component of the infrastructure of modern urban areas \cite{UPTmodes}. These mass transportation systems contribute to solving the problem of connecting people with different parts of a city, especially in densely populated urban areas, allowing them to carry out the diverse activities that keep a city functioning \cite{Ortuzar, su12176844,Derrible2012}.  Under optimal conditions, metro systems can move large numbers of people in a fast and efficient way,  with lower transportation costs and coping with traffic-related air pollution  \cite{Leong,pollution,To_Lee_Yu_2020}.  Thereby metro systems have become a crucial element in the development of cities not only in terms of economical benefits but also in terms of social and environmental impact. 
\\[2mm]
Unfortunately, metro systems are exposed to a  considerable number of factors that can compromise their operation.  Lack of maintenance, technical disruptions, natural events, are some of the issues that affect their correct functioning, and in some cases, they can pose a risk for passengers \cite{Alireza,Gao_Wang_2021}. Therefore it is of crucial importance understanding these systems and their response to damage. Several works have been developed around metro systems, particularly as they represent real networks, many of them  focused on network science \cite{Derrible2012, Derrible_Kennedy_2010}. In this representation, the components of metro systems are described as nodes of a graph and their relationships as edges connecting them.  Under this approach, it has been observed that metro systems can exhibit properties like scale-free and small-world features \cite{Derrible2012,Derrible_Kennedy_2010}.
The response of these systems to damage has been studied in this context too, principally using topological aspects of networks. Generally, the description of failures in networks has been modeled based on processes that involve the removal of sets of nodes or links through random, targeted or cascade-base attacks \cite{Boccaletti_Latora_Moreno_Chavez_Hwang_2006,Cohen_Havlin_2010,Zhang_Fu_Li_2016}, that in a metro system represents the complete dysfunction of some stations, rails, sections of a road, among others \cite{Berche_von_Ferber_Holovatch_Holovatch_2009,Berche_von_Ferber_Holovatch_Holovatch_2010, Ferber_Berche_Holovatch_Holovatch_2012}. In the same way, the performance in some metro systems has been assessed based on elements of network theory such as percolation methods, measures of connectivity as betweenness centrality, optimal and redundant paths and effective graph resistance \cite{Derrible_Kennedy_2010b, wang2015quantifying, Wang_Pournaras_Kooij_VanMieghem_2014}. In addition, partial failures of  links and nodes have been used as a model of damage in transportation problems as well, although these cases are more common in the real world \cite{Cats_Jenelius_2018,Cats_Koppenol_Warnier_2017,Ye_Kim_2019} they have been studied to a lesser extent \cite{Pan_Yan_He_He_2021}.  Scenarios of partial failures  occur when some components of the network do not suffer complete dysfunction but its performance is affected so that  the service capacity, travel demands, delays, may change \cite{Cats_Jenelius_2018,Cats_Koppenol_Warnier_2017,Ye_Kim_2019}. 
\\[2mm]
On the other hand, the  understanding of different dynamical processes on networks has had a significant impact \cite{VespiBook}. In particular, the diffusive transport described by a random walker that visits the nodes on networks following different rules is a challenging theoretical problem where one of the main goals is to understand the relation between network topology and its capacity to communicate all the nodes in the network \cite{Hughes,MasudaPhysRep2017,FractionalBook2019,RiascosJCN2021}. Different developments in the understanding of random walkers on networks have led to valuable tools in searching processes on the internet \cite{Brin1998,ShepelyanskyRevModPhys2015}, algorithms for data mining \cite{BlanchardBook2011, LeskovecBook2014},  human mobility in cities \cite{Barbosa2018,RiascosMateosPlosOne2017,RiascosMateosSciRep2020}, epidemic spreading \cite{Bestehorn2021}, among many others.
\\[2mm]
In this contribution, we explore the effect of damage impact in metro systems considering that the main function of these systems is to maintain the capacity that an agent can efficiently reach any node from any initial condition. In this manner, it is reasonable to describe the transport in the structure in terms of random walkers defined using local information of the links connecting two nodes. For such dynamical processes, a global time defined in terms of mean first passage times between nodes is a useful quantity to describe the functionality of the system in terms of eigenvalues and eigenvectors of the transition matrix that defines the dynamics. In addition, at a different scale of time, we consider that the transportation network may suffer damage in the links, where this degradation in the link capacity is modeled by a stochastic process with preferential attachment in such a way that links with more failures are susceptible to new deficiencies with greater probability. Our modeling is a proxy of the response of the infrastructure of a metro system under situations that do not represent extreme events but with an accumulation of partial failures. We study the evolution of the functionality of metro systems under this particular mechanism for accumulation of damage. This approach allows us to characterize the response to link deterioration and classify their structures from vulnerable to robust under failure. This classification also relates the topology of the network with its capacity to tolerate damage. The application explored in this paper occurs in the context of the analysis of infrastructure in urban transportation systems; however, the methods presented are general and can be implemented to analyze the vulnerability of different `complex' systems.

\section{Dataset description}
\label{DataSet}
The metro system is one of the mass transportation systems used in several urban areas worldwide. Its infrastructure constitutes a rail system with exclusive right-of-way whose tracks can be underground, at grade, or elevated. It works based on previously designed routes or lines connecting several parts of the city and, usually, it is integrated to other transportation modes; nevertheless, its operation is independent of them \cite{UPTmodes}.
\\[2mm]
There are several ways to represent a transportation system as a network that can provide different information about its structure \cite{von2009}. For instance, the $\mathbb{L}$-space representation considers each station as a node of the graph, and two of them are linked only if there is a route that connects them directly \cite{von2009}. In another representation, the $\mathbb{B}$-space, the routes are considered nodes, the edges are drawn between the routes and the stations that belong to their path, thus the stations appear connected through routes only \cite{von2009}. In this work, we use a representation proposed by Derrible \cite{Derrible2012}. In this depiction of a metro system, the vertices correspond only to the terminal stations which are the stations at the end of a line and the transfer stations that allow people to change from one line to another. Other stations are ruled out since they do not provide relevant information. To draw the edges, Derrible considers two particular types: in the first group of edges are the simple ones defined by the connections that establish a single line without overlapping. The second group considers multiple edges representing the fact that there is more than a line linking the stations. In the following, we dismiss the effects of overlapping lines, thereby the metro systems are represented by networks of terminal and transfer stations connected by directed edges.    
\\[2mm]
\begin{figure*}[t!]
	\begin{center}
		\includegraphics*[width=1\textwidth]{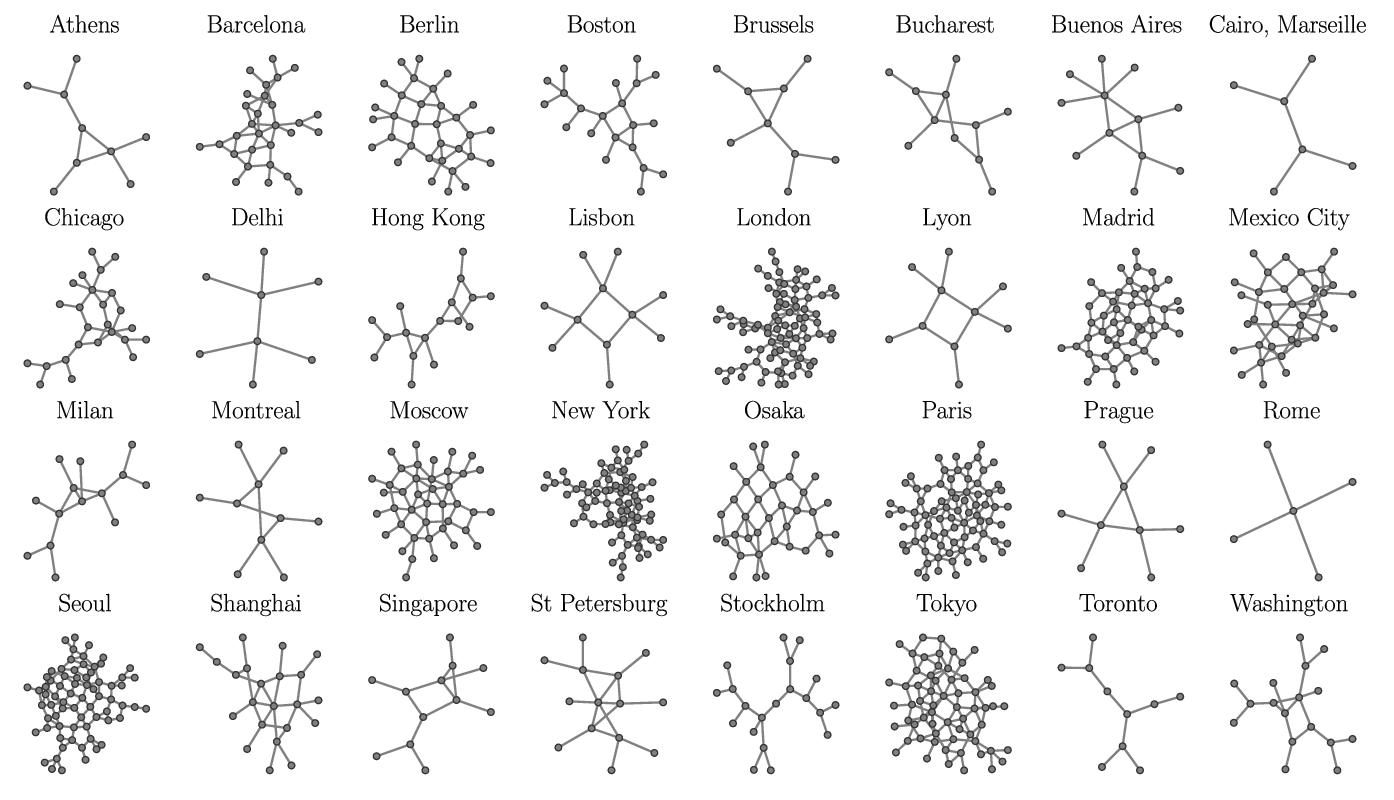}
	\end{center}
	\vspace{-5mm}
	\caption{\label{Fig_1} Networks of 33 metro systems worldwide. The adjacency matrices of the networks were obtained from Ref. \cite{Derrible2012,SybilWolfram2014} and plotted using the networkx (2.6.3) python package \cite{NetworkX}.}
\end{figure*}
We study  33 metro systems using the information and database compiled by Derrible \cite{Derrible2012,SybilWolfram2014} and depicted in Fig. \ref{Fig_1}. It is important to mention this dataset for the metro systems was collected in 2008–2009, and discrepancies to current infrastructure might exist (this is the case of the metro in Rome where a couple of lines have been added or the system in Shanghai that has substantially increased). We use the dataset in Ref. \cite{Derrible2012,SybilWolfram2014} due to all the documentation describing the methods to collect the data and the results obtained with standard metrics of network science (see details in Refs. \cite{Derrible2012,DerribleThesis2010,SybilWolfram2014}). 
The systems analyzed belong to different parts of the world (North and Latin America, Eastern and Western Europe, Africa and Asia) and have diverse topologies.  Regarding the number of lines, they range from 2 lines in the Cairo system to 14 lines found in the metro of Paris \cite{Derrible2012}. For the number of nodes, using the representations of terminals and transfer stations, we can find small graphs as the Rome metro ($N=5$) or Cairo and Marseille systems ($N=6$) and large ones as the case of  New York ($N=77$) or London ($N=83$). Referring to the structure, the data set contains systems represented by connected networks with a tree structure, for instance, the case of Cairo, Delhi, or Toronto.   Other systems show a few cycles in their structures as in Prague or Bucharest and some of them display more complex structures such as the systems in Paris or London.
\section{Transportation networks under cumulative damage}
This section summarizes the methods introduced in Refs.  \cite{Aging_PhysRevE2019,Eraso_Hernandez_2021} to the study of random walks on systems that evolve with accumulation of damage. For the initial configuration, we consider undirected connected networks with $N$ nodes $i=1,\ldots ,N$ described by an adjacency matrix $\mathbf{A}$ with elements $A_{ij}=A_{ji}=1$ if there is an edge between the nodes $i$ and $j$ and $A_{ij}=0$ otherwise; in particular, $A_{ii}=0$ to avoid edges connecting a node with itself. In this structure, we denote the set of nodes as $\mathcal{V}$ and the set of directed edges as $\mathcal{E}$ with elements $(i,j)$, $|\mathcal{E}|$ is the total number of different directed edges in the network.
\\[2mm]
Additionally to the network structure, the global state of the system at time $T=0,1,2,\ldots$ is characterized by a $N\times N$ matrix of weights $\mathbf{\Omega}(T)$ with elements $\Omega_{ij}(T)\geq 0$ and $\Omega_{ii}(T)=0$ which describe weighted connections between the nodes. The matrix $\mathbf{\Omega}(T)$ contains information of the state of the edges, in general is not symmetric and contains the effect of damage at time $T$. We introduce for each link $(i,j) \in \mathcal{E}$ a stochastic integer variable $h_{ij}(T)$ where $h_{ij}(T)-1$ counts the number of random faults that exist in this link at time $T$. The values $h_{ij}(T)$ for all the edges are numbers that evolve randomly, and a new fault in the link $(i,j)$ appears at time $T$ with a probability $\pi_{ij}(T)$ which is given by
\begin{equation}\label{problinks}
	\pi_{ij}(T)=\frac{h_{ij}(T-1)}{\sum_{(l,m) \in \mathcal{E}} h_{lm}(T-1)}\qquad (i,j) \in \mathcal{E},
\end{equation}
for $\,T=1,2,\ldots$ with the initial condition $h_{ij}(0)=1$, i.e. no faults exist for all the edges at $T=0$. For the sake of an undamaged reference edge during the damage evolution we choose randomly an edge $\mathcal{E}^\star$ which does not evolve according to Eq. (\ref{problinks}) and maintains $h_{\mathcal{E}^\star}(T)=1$, i.e. remains without damage for all $T \geq 0$.
\\[2mm]
Equation (\ref{problinks}) indicates the probability for the event that at time $T$ the number of faults $h_{ij}(T)=h_{ij}(T-1)+1$ is increased by one. In this manner, the variable $T$ is a measure of the total number of hits in the links of the network.
\\[2mm]
In our analysis, the damage is distributed without maintaining the symmetry of the initially undirected network, i.e. damage in the edge $(i,j)$ evolves independently of
the damage in edge $(j,i)$ thus in the general case $h_{ij}(T)$ is independent of the value $h_{ji}(T)$ and also $\pi_{ij}(T)$ from $\pi_{ji}(T)$ which is generating a biased network. 
With Eq. (\ref{problinks}) at $T=1$ the first hit (fault) is randomly generated for any selected link $(i,j)$
with equal probability $\pi_{ij}(1)=\frac{1}{|\mathcal{E}|-1}$ for $(i,j)\in \mathcal{E}\setminus\mathcal{E}^\star$ (where $\mathcal{E}\setminus\mathcal{E}^\star$ denotes the set of edges minus the particular edge $\mathcal{E}^\star$). The occurrence of the second fault at $T=2$ depends on the previous configuration and so on. 
\\[2mm]
An essential feature of the probabilities in Eq. (\ref{problinks}) is that they produce preferential damage if a link has already suffered damage in the past. A link has a higher probability to get a fault with respect to a link never being damaged. Such preferential random processes have been explored in different contexts in science (see Ref. \cite{NetworkScienceBook2016}), being a key element in our model that generates complexity in the distribution of damage reflected by asymptotically emerging power-law and fractal features. An asymptotic analysis of the time-evolution of the fault number distribution resulting from Eq. (\ref{problinks})
shows that a power-law scaling with features of a stochastic fractal emerge (see Ref. \cite{Aging_PhysRevE2019}).
Such a preferential 
damage accumulation mechanism can be observed in several adaptive complex systems such as living beings and was suggested as a model for aging  \cite{WangMiWun2009,WangMi2015,WangMi2015b}.
\\[3mm]
Now, we aim to describe how the structure reacts to the damage hits occurring stochastically to the edges. We describe the effects of the damage by using the information in the matrix of weights $\mathbf{\Omega}(T)$. In terms of the values $h_{ij}(T)$, the matrix $\mathbf{\Omega}(T)$ defines the global state of the 
network containing the complete information on the network topology at time $T$.
Its matrix elements 
\begin{equation}\label{OmegaijT}
	\Omega_{ij}(T)=(h_{ij}(T))^{-\alpha} A_{ij}
\end{equation}
contain the local information on the damaged state of edge $(i,j)$
and $\alpha\geq 0$ is a real-valued parameter that quantifies the effect of the damage in each link. The  parameter $\alpha$ describes the reaction of the system to the damage in the links and can be conceived as a `maintenance parameter'. It quantifies activities of maintenance and reparation, for instance small $\alpha$ corresponds to well maintained and therefore robust metro networks and large $\alpha$ reflect metro systems which are fragile due to bad maintenance (see Refs. \cite{WangMiWun2009,WangMi2015,WangMi2015b} for a discussion of the analogue so-called `misrepair' mechanism in living beings). In the limit $\alpha\to 0$ we have $\Omega_{ij}(T) \to A_{ij}$ as in a perfect undamaged structure (perfect maintenance), and the effect of the stochastically generated faults is null. In contrast, in the limit $\alpha\to\infty$, a hit in a link is equivalent to its removal from the network (no maintenance).  In the following, we use a finite value of $\alpha$ to deal with scenarios before reaching extreme damage, we center our discussion on the evolution due to cumulative damage.
\\[2mm]
In addition to the damage accumulation of the structure characterized by $T$, at a completely different scale of times $t$ (significantly less than the characteristic times of damage evolution, i.e, $\Delta t\ll \Delta T$) takes place the movement of random walkers in the network with discrete steps at times $t=0,\Delta t, 2\Delta t,\ldots$. This time scale can be identified with the characteristic operational time of a metro train, for instance, the time between two stations which may be in the order of magnitude of minutes to hours whereas the degradation of a metro line may take months or even years.
In a determined configuration at time $T$, the transition probability matrix $\mathbf{W}(T)$ describing the random walker is defined by the elements $w_{i\to j}(T)$ with the probability to pass from node $i$ to node $j$
\begin{equation}\label{transitionPij}
	w_{i\to j}(T)=\frac{\Omega_{ij}(T)}{\sum_{\ell=1}^N\Omega_{i\ell}(T)}.
\end{equation}
We assume a {\it Markovian} 
time-discrete random walker that performs at any time increment $\Delta t$ a random step from one node to another.
This process is defined by the master equation \cite{Hughes,LambiottePRE2011,NohRieger2004}
\begin{equation}
	\label{mastereqnNRW}
	P_{ij}(t+\Delta t,T) = \sum_{\ell=1}^NP_{i\ell}(t,T)w_{\ell\to j}(T) 
\end{equation}
valid at the small scale of times $t$ for which we assume  $\mathbf{W}(T)$ is constant allowing us to characterize the dynamics of the random walker defined by $\mathbf{W}(T)$ at $T=0,1,2,\ldots$. In this master equation $P_{ij}(t,T)$ indicates the probability that the walker that starts its walk at node $i$ at $t=0$ occupies node $j$ at the $n$-th time step $t=n\Delta t$ (in the following the times characterizing the random walk dynamics are expressed as multiples of $\Delta t$ and $\Delta t\ll \Delta T=1$).
\\[2mm]
We study the capacity of the network to perform a specific function and how this property evolves with the accumulation of damage. In the context of transport on networks, we use a `functionality' $\mathcal{F}(T)$ that quantifies the global transport capacity at time $T$ as \cite{Aging_PhysRevE2019}
\begin{equation}\label{F_ratioT}
	\mathcal{F}(T)\equiv\frac{\tau(0)}{\tau(T)}
\end{equation}
with
\begin{equation}
	\label{globaltime_tau}
	\tau(T)= \frac{1}{N}\sum_{j=1}^N \tau_j(T),
\end{equation}
where
\begin{equation}\label{TauiSpect}
	\tau_j(T)=\sum_{l=2}^N\frac{1}{1-\lambda_l(T)}\frac{\left\langle j|\phi_l(T)\right\rangle \left\langle\bar{\phi}_l(T)|j\right\rangle}{\left\langle j|\phi_1(T)\right\rangle \left\langle\bar{\phi}_1(T)|j\right\rangle}\, .
\end{equation}
Here, we use Dirac's  notation, $|\phi_m(T)\rangle, \langle {\bar \phi}_m(T)|$ denote, respectively, the right and left  eigenvectors of the transition matrix  $\mathbf{W}(T)$ with the respective eigenvalues $0\leq |\lambda_m(T)| \leq 1$. The walk which we assume to take place on a (strongly connected) directed weighted and finite network is ergodic with the unique eigenvalue $\lambda_1(T) =1 \forall T $. The stationary distribution $P_{j}^{(\infty)}(T)$, that gives the probability to find the random walker at the node $j$ in the limit $t\to\infty$, is given by $P_{j}^{(\infty)}(T) = \langle i|\phi_1(T)\rangle\langle \bar \phi_1(T)|j\rangle$ \cite{MasudaPhysRep2017,FractionalBook2019,RiascosJCN2021,RiascosMateos2012}. The global time $\tau(T)$ expressed in terms of mean first passage  time $\left\langle {\cal T}_{ij}(T)\right\rangle$ to start in $i$ and reach for the first time the node $j$ is given by \cite{Eraso_Hernandez_2021}
\begin{equation}\label{TauGlobalMFPTdef}
	\tau(T)=\frac{1}{N}\sum_{j=1}^N\sum_{i\neq j}P_{i}^{(\infty)}\left\langle {\cal T}_{ij}(T)\right\rangle.
\end{equation}
In this result, we see that $\tau(T)$ is a global time that gives the weighted average of the number of steps to reach any node. In this case, each $\left\langle {\cal T}_{ij}(T)\right\rangle$ is weighted with the value $P_{i}^{(\infty)}$ that acts as a ranking of the nodes in the dynamics defined by each matrix $\mathbf{W}(T)$.  In this way, the definition  $\mathcal{F}(T)$ in Eq. (\ref{F_ratioT}) characterizes globally the effect of the damage suffered by the whole structure and how evolves the capacity of a random walker to explore the network. For $\alpha>0$, the smaller $\tau(T)$ (i.e. the higher the transport capacity),
the higher the functionality. For large times $T\gg 1$, the value $\tau(T) \geq \tau(0)$, therefore $\mathcal{F}(T) \leq 1$
(equality holds only in the undamaged state) \cite{Aging_PhysRevE2019}. Here, it is important to mention that additionally to the damage evolution in Eq. (\ref{problinks}), we have introduced the condition that $h_{\mathcal{E}^\star}(T)=1$ is kept constant for a randomly chosen edge $\mathcal{E}^\star$. This particular restriction is necessary to maintain the link $\mathcal{E}^\star$ without damage as a reference of the complete functionality of a link and to avoid sudden ``revival'' of the system described by the matrix $\mathbf{W}(T)$ which may randomly occur if all the links have suffered at least one fault, see Ref.  \cite{Aging_PhysRevE2019} for a detailed discussion.
\section{Metro systems under damage}
\begin{figure*}[t!]
	\begin{center}
		\includegraphics*[width=1.0\textwidth]{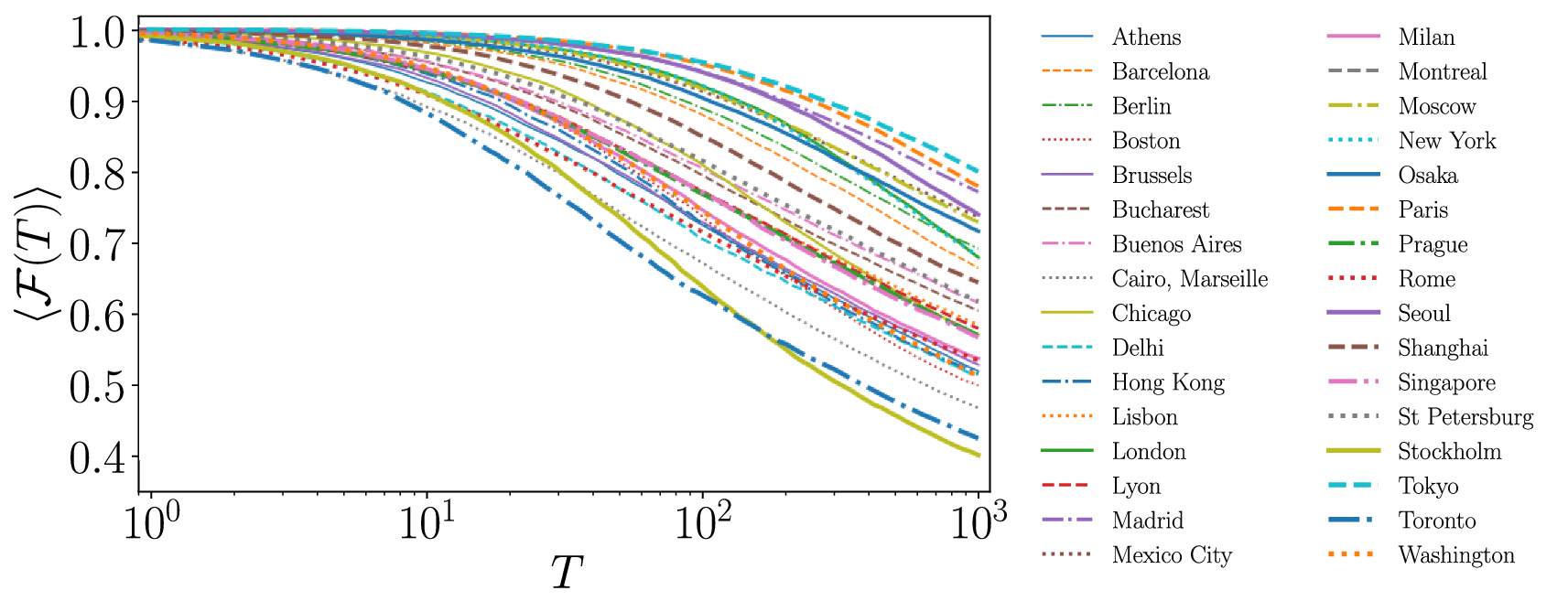}
	\end{center}
	\vspace{-5mm}
	\caption{\label{Fig_3} Evolution of the ensemble average $\langle\mathcal{F}(T)\rangle$ for 33 metro systems worldwide. Each curve is obtained with the average of 1000 Monte Carlo realizations of the cumulative damage algorithm with $\alpha=0.5$ for $1\leq T\leq 10^3$.}
\end{figure*}

Once we have defined the algorithm for the accumulation of damage in transportation networks, in this section we apply this approach to understand and classify the metro systems presented in Sec. \ref{DataSet}. To this end, we implement Monte Carlo simulations to generate random distributions of damage $h_{ij}(T)$ in the links at times $T=1,2,\ldots,T^\star$ by using Eq. (\ref{problinks}). For each network and a  distribution of damage at time $T$, we build a transition matrix $\mathbf{W}(T)$ using Eqs. (\ref{OmegaijT})-(\ref{transitionPij}) describing a random walker in the structure with damage. The eigenvalues and eigenvectors of this matrix allow to characterize the capacity of transport in terms of $\tau(T)$ in Eq. (\ref{globaltime_tau}) and the functionality $\mathcal{F}(T)$ in Eq. (\ref{F_ratioT}). Although we are working with an idealization of how accumulated damage occurs in these structures, all this formalism in the context of random walkers allows us to quantify the resistance of a network dedicated to transport, its ability to tolerate damage as well as the relationship between the network topology and its robustness. The importance of the network topology can be seen by the simple observation that if an edge between two nodes is strongly damaged or disconnected, then the functionality of the structure is not much reduced if redundant short paths connecting these two nodes exist.
\\[2mm]
%
\begin{figure}[!t]
	\begin{center}
		\includegraphics*[width=0.7\textwidth]{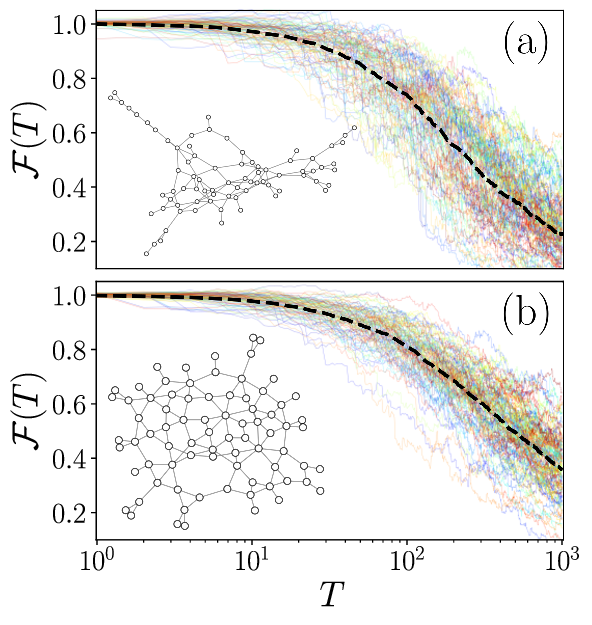}
	\end{center}
	\vspace{-5mm}
	\caption{\label{Fig_2} Evolution of $\mathcal{F}(T)$ for two metro systems in: (a) New York  and (b) Paris. The results for 100 Monte Carlo realizations of the cumulative damage algorithm with $\alpha=1$ are presented with thick lines and dashed lines depict the ensemble average $\langle\mathcal{F}(T)\rangle$. In each panel we include the network analyzed.}
\end{figure}
%
In Fig. \ref{Fig_2} we illustrate the results for the functionality $\mathcal{F}(T)$ as a function of $T$ (in the interval $T\in [1,1000]$) for two particular metro systems in New York  [Fig. \ref{Fig_2}(a)] and Paris [Fig. \ref{Fig_2}(b)] using $\alpha=1$ in Eq. (\ref{OmegaijT}). In the different curves, we depict 100 realizations where we can see how $\mathcal{F}(T)$ changes  as the systems receive more damage. The values $\langle\mathcal{F}(T)\rangle$ in dashed lines represent the ensemble average.
The systems explored present two different topologies as we can see in the insets. For the case of New York, the network explored contains $N=77$ nodes and $|\mathcal{E}|=218$ edges. In the transportation network in Paris we have $N=78$ and $|\mathcal{E}|=250$. 
\\[2mm]
In Fig. \ref{Fig_2}, it is worth mentioning that by definition we have $\mathcal{F}(0)=1$. The ensemble average $\langle\mathcal{F}(T)\rangle$ decreases monotonically in the interval explored showing that in the transportation networks the common effect produced by the damage is the increasing of the times to explore the network, i.e. $\tau(T)>\tau(0)$, in other words a damaged metro system increases the traveling time of the passengers. The results also reveal cases where $\mathcal{F}(T)>1$, this effect could appear when there is at the periphery of the system a subnetwork with a small set of nodes acting like a ``trap’’ where the random walker spends much time and where it is hard to escape to continue the global exploration of the structure.  Then, some damage in the links reducing the probability to visit this part of the network may be beneficial, reducing global exploration times.
\\[2mm]
Although the most common behavior is the reduction of functionality with damage, in some Monte Carlo realizations it can be seen how the action of the damage improves the transport capacity with respect to the previous configuration; for example, creating a local bias that is more effective (see Ref. \cite{DirectedFractional_PRE2020} for a discussion on the effect of bias in ergodic random walks). In these cases $ \tau(T+1)<\tau(T)$, therefore $ \mathcal{F}(T + 1) >\mathcal {F} (T) $. However, in the ensemble average we see that $\langle\mathcal{F}(T+1)\rangle<\langle\mathcal{F}(T)\rangle$. On the other hand, the simulations show that the damage affects the New York network more quickly than the Paris network. In general, in New York $\langle \mathcal {F} (T)\rangle $ decays faster with $T $. This effect may be due to the existence of fewer links in the New York network. Nevertheless, as we will see in the following part, the number of links is only one of the factors that modify the capacity of the structure to tolerate damage.
\\[2mm]
Let us now apply the same method implemented for New York and Paris in Fig. \ref{Fig_2} to the 33 metro systems in Fig. \ref{Fig_1}. In Fig. \ref{Fig_3} we present the results for the ensemble average $\langle\mathcal{F}(T)\rangle$ as a function of $T$ for values in the interval $1\leq T \leq 10^3$ considering $1000$ Monte Carlo simulations of the process with cumulative damage with $\alpha=0.5$. The results show how $\langle\mathcal{F}(T)\rangle$ evolves with the damage in the different systems. 
Each curve $\langle\mathcal{F}(T)\rangle$ is a characterization of the system that includes its capacity to transport and the response of the whole structure under damage. In some networks, we see a fast decay of $\langle\mathcal{F}(T)\rangle$ (see for example the cases of Toronto and Stockholm) revealing that a few hits in the edges reduce significantly the communicability of the system. In contrast, other networks have a redundant structure (Paris, Tokyo) and can tolerate the damage, a fact that is described with the slow decay of the values  $\langle\mathcal{F}(T)\rangle$.
\section{Normalized cumulative damage}
In addition to our discussion about the evolution of the ensemble average $\langle\mathcal{F}(T)\rangle$, the gradual reduction of this quantity can be associated with the ability of a system to operate under damage and with its robustness. We are interested in comparing the metro systems to determine which network structure is the most robust under accumulation of damage. However, here it is important to notice as a consequence of damage occurring in the edges, more links can generate apparently greater resistance by the existence of redundant short paths connecting some pairs of nodes. Furthermore, all the infrastructures associated to a link also mean a cost in the initial configuration of the system. Therefore, it is more pertinent to normalize $\langle\mathcal{F}(T)\rangle$ using the quantity \cite{Eraso_Hernandez_2021}
\begin{equation}
	\rho_\mathcal{E}=\frac{|\mathcal{E}|}{N(N-1)},
\end{equation}
where $|\mathcal{E}|=\sum_{l,m=1}^N A_{lm}$ is the total number of edges (including the direction of each link) and $N(N-1)$ is the total number of connections on a fully connected graph without loops and we have $\rho_\mathcal{E}\leq 1$. Networks with  $\rho_\mathcal{E}\approx 1$ are densely connected whereas $\rho_\mathcal{E}\ll 1$ are typical for networks with few edges, this is the case of trees and rings.
\\[2mm]
\begin{figure*}[t!]
	\begin{center}
		\includegraphics*[width=1.0\textwidth]{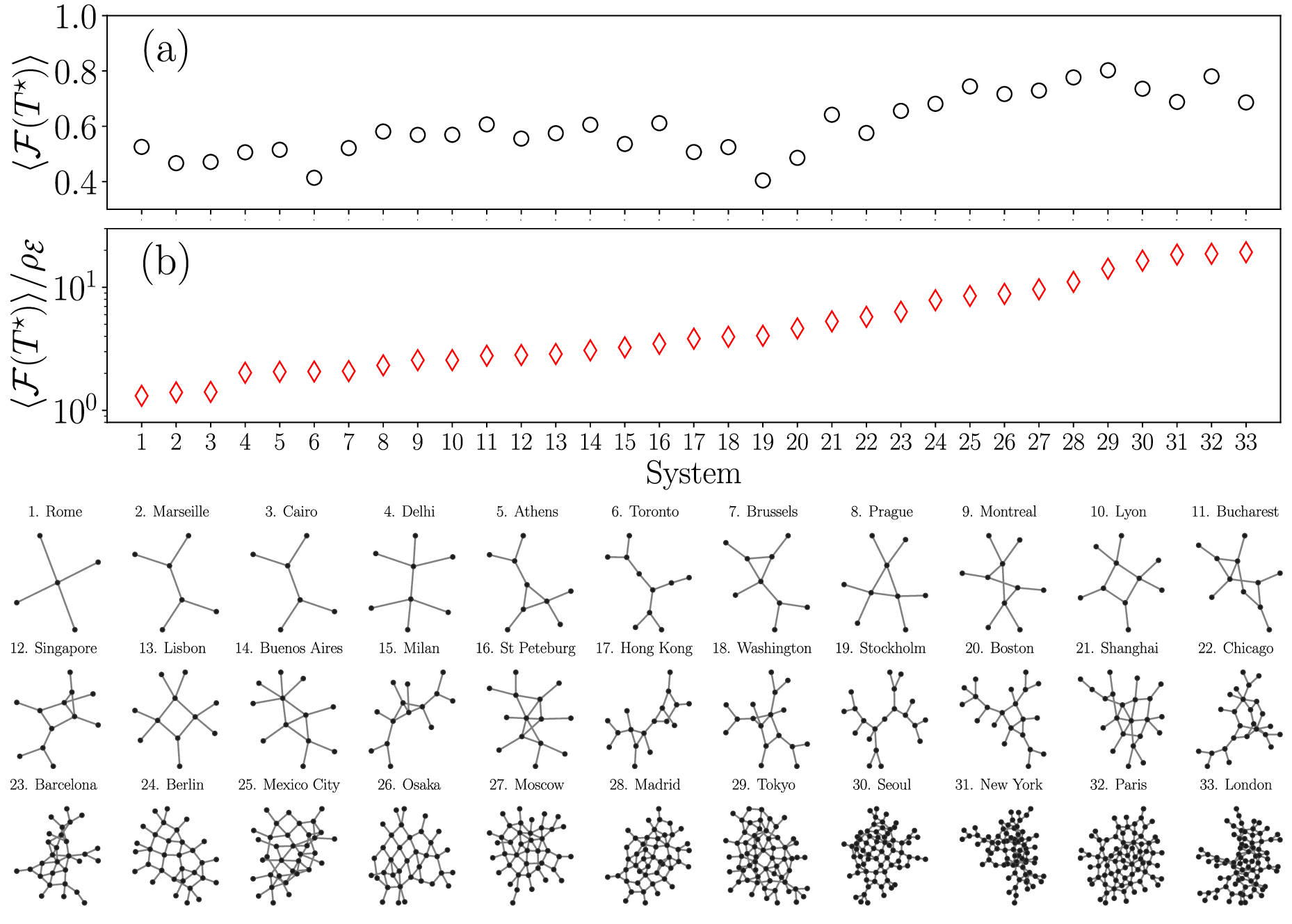}
	\end{center}
	\vspace{-5mm}
	\caption{\label{Fig_4} Ensemble average of the functionality at $T^\star=10^3$ for the transport on different network metro systems with cumulative damage generated with $\alpha=0.5$ and 1000 realizations. We present (a) $\langle\mathcal{F}(T^\star)\rangle$  and (b) the normalized value $\langle\mathcal{F}(T^\star)\rangle/\rho_{\mathcal{E}}$.}
\end{figure*}
In Fig. \ref{Fig_4}(a) we present the values of $\langle\mathcal{F}(T^\star)\rangle$ at time $T^\star=10^3$, $\alpha=0.5$ using 1000 realizations. In Fig. \ref{Fig_4}(b) we show the normalized values $\langle\mathcal{F}(T^\star)\rangle/\rho_{\mathcal{E}}$. In both cases, the networks are sorted using the values $\langle\mathcal{F}(T^\star)\rangle/\rho_{\mathcal{E}}$, systems are numbered from 1 to 33 and presented at the bottom of panel (b).
\\[2mm]
\begin{table}[!t]
	\centering
		\tbl{\label{Table1} Characterization of the 33 metro systems in Fig. \ref{Fig_4}. For each network we present the number nodes $N$, the total number of edges $|\mathcal{E}|$, the average degree $\bar{k}$, the mean triangle clustering $\mathcal{C}_\triangle$, the mean  clustering $\mathcal{C}_\square$ associated to squares, the global time $\tau(0)$  and the normalized functionality $\langle\mathcal{F}\rangle/\rho_{\mathcal{E}}$ with $\langle\mathcal{F}\rangle=\langle\mathcal{F}(T^\star)\rangle$ evaluated at $T^\star=10^3$ using the ensemble average with $1000$ realizations and $\alpha=0.5$.}{
	\begin{tabular}{l c c c cc c c}
		\hline    
		{\bf System} & $N$ &  $|\mathcal{E}|$ & $\bar{k}$ & $\mathcal{C}_\triangle$ & $\mathcal{C}_\square$ & $\tau(0)$ & $\langle\mathcal{F}\rangle/\rho_{\mathcal{E}}$  \\[0.5mm]
		\hline  
		1. Rome 	 & 5 &  8 &  1.6 &  0.0 &  0.0 &  5.3 &  1.31 \\
		2. Marseille 	 & 6 &  10 &  1.67 &  0.0 &  0.0 &  8.2 &  1.4 \\
		3. Cairo 	 & 6 &  10 &  1.67 &  0.0 &  0.0 &  8.2 &  1.41 \\
		4. Delhi 	 & 8 &  14 &  1.75 &  0.0 &  0.0 &  12.9 &  2.02 \\
		5. Athens 	 & 9 &  18 &  2.0 &  0.09 &  0.0 &  17.1 &  2.06 \\
		6. Toronto 	 & 10 &  18 &  1.8 &  0.0 &  0.0 &  23.1 &  2.07 \\
		7. Brussels 	 & 9 &  18 &  2.0 &  0.09 &  0.0 &  16.8 &  2.09 \\
		8. Prague 	 & 9 &  18 &  2.0 &  0.06 &  0.0 &  15.0 &  2.32 \\
		9. Montreal 	 & 10 &  20 &  2.0 &  0.0 &  0.04 &  17.3 &  2.56 \\
		10. Lyon 	 & 10 &  20 &  2.0 &  0.0 &  0.04 &  17.3 &  2.56 \\
		11. Bucharest 	 & 11 &  24 &  2.18 &  0.06 &  0.0 &  19.8 &  2.78 \\
		12. Singapore 	 & 12 &  26 &  2.17 &  0.06 &  0.02 &  25.7 &  2.82 \\
		13. Lisbon 	 & 11 &  22 &  2.0 &  0.0 &  0.03 &  19.9 &  2.87 \\
		14. Buenos Aires 	 & 12 &  26 &  2.17 &  0.07 &  0.01 &  23.0 &  3.07 \\
		15. Milan 	 & 14 &  30 &  2.14 &  0.07 &  0.01 &  34.5 &  3.25 \\
		16. St Petersburg 	 & 14 &  32 &  2.29 &  0.07 &  0.01 &  29.1 &  3.48 \\
		17. Hong Kong 	 & 17 &  36 &  2.12 &  0.04 &  0.0 &  48.3 &  3.83 \\
		18. Washington 	 & 17 &  36 &  2.12 &  0.04 &  0.02 &  45.5 &  3.96 \\
		19. Stockholm 	 & 20 &  38 &  1.9 &  0.0 &  0.0 &  72.7 &  4.04 \\
		20. Boston 	 & 21 &  44 &  2.1 &  0.03 &  0.01 &  65.5 &  4.64 \\
		21. Shanghai 	 & 22 &  56 &  2.55 &  0.05 &  0.05 &  54.8 &  5.29 \\
		22. Chicago 	 & 25 &  60 &  2.4 &  0.07 &  0.02 &  75.7 &  5.76 \\
		23. Barcelona 	 & 29 &  84 &  2.9 &  0.17 &  0.03 &  85.3 &  6.34 \\
		24. Berlin 	 & 32 &  86 &  2.69 &  0.08 &  0.02 &  84.0 &  7.86 \\
		25. Mexico City 	 & 35 &  104 &  2.97 &  0.1 &  0.04 &  84.3 &  8.51 \\
		26. Osaka 	 & 36 &  102 &  2.83 &  0.08 &  0.03 &  94.4 &  8.85 \\
		27. Moscow 	 & 41 &  124 &  3.02 &  0.09 &  0.04 &  113.1 &  9.64 \\
		28. Madrid 	 & 48 &  158 &  3.29 &  0.13 &  0.05 &  126.9 &  11.09 \\
		29. Tokyo 	 & 62 &  214 &  3.45 &  0.15 &  0.03 &  167.0 &  14.18 \\
		30. Seoul 	 & 71 &  222 &  3.13 &  0.09 &  0.04 &  234.3 &  16.47 \\
		31. New York 	 & 77 &  218 &  2.83 &  0.05 &  0.02 &  280.0 &  18.47 \\
		32. Paris 	 & 78 &  250 &  3.21 &  0.13 &  0.02 &  239.2 &  18.75 \\
		33. London 	 & 83 &  242 &  2.92 &  0.1 &  0.02 &  329.1 &  19.3 \\[2mm]
		\hline
	\end{tabular}
}
\end{table}
The networks as sorted in Fig. \ref{Fig_4} give us a numerical and graphical representation of the existing connection between the robustness of the network and its topology. We observe that systems with low $\langle\mathcal{F}(T^\star)\rangle/\rho_{\mathcal{E}}\leq 2.78$ are extremely fragile and just one hit can reduce abruptly the functionality; networks with this characteristic have a tree structure and few nodes (systems 1-4, 6) the incorporation of one cycle with three nodes (systems 5, 7, 8) or four nodes (as in systems 9 to 10) increases the tolerance to damage. More robust systems (systems 11 to 22) have diverse proportions of cycles with different lengths. In this manner, the degradation of the functionality of a link can be compensated with the existence of multiple paths connecting two nodes maintaining operational conditions for the global transport. In this classification of the metro structures, systems 23 to 33 have the strongest topologies with $\langle\mathcal{F}(T^\star)\rangle/\rho_{\mathcal{E}}>6.0$.
\\[2mm]
We complement the characterization of metro systems  with the numerical values in Table \ref{Table1}. In this table we report different types of quantities for the systems as shown in Fig. \ref{Fig_4}. We include the number of nodes $N$, the number of edges $|\mathcal{E}|$, and the average degree $\bar{k}=\frac{1}{N}\sum_{i=1}^N k_i$ (with $k_i=\sum_{l=1}^N A_{il}$), these quantities give us a first description of the network. However, in our analysis it is necessary to  delve into the connectivity of the network; in particular, the proportion of cycles with three (triangles) and four nodes (squares). To this end we explore the clustering coefficient $\mathcal{C}_3(i)$ of the node $i$ that quantifies the fraction of connected neighbors ${\triangle}_i$ (triangles) of  the node $i$ with respect to the maximum number of these connections given by $k_i(k_i-1)/2$. In terms of the adjacency matrix we have for $k_i\geq 2$ (see
 \cite{NewmanBook} for details)
\begin{equation}
	\mathcal{C}_3(i)=\frac{ (\mathbf{A}^3)_{ii}}{k_i(k_i-1)},
\end{equation}
otherwise $\mathcal{C}_3(i)=0$. The global average coefficient is given by 
\begin{equation}
	\mathcal{C}_\triangle=\frac{1}{N}\sum_{i=1}^N \mathcal{C}_3(i).
\end{equation}
In a similar way, we can evaluate the fraction of squares that exists at the node $i$ by using
\begin{equation}
	C_4(i) = \frac{ \sum_{l=1}^{k_i} 
		\sum_{m=l+1}^{k_i} q_i(l,m) }{ \sum_{l=1}^{k_i} 
		\sum_{m=l+1}^{k_i} [a_i(l,m) + q_i(l,m)]},
\end{equation}
where $q_i(l,m)$ are the number of common neighbors of $l$ and $m$ 
other than $i$ (i.e. squares), and 
$a_i(l,m) = (k_l - (1+q_i(l,m)+\theta_{lm}))+(k_m - (1+q_i(l,m)+\theta_{lm}))$,
where $\theta_{lm} = 1$ if $l$ and $m$ are connected and 0 otherwise (see Ref. \cite{Zhang_PhysicaA_2008} for a detailed discussion of these quantities). In terms of $\mathcal{C}_4(i)$, we define 
\begin{equation}
	\mathcal{C}_\square=\frac{1}{N}\sum_{i=1}^N \mathcal{C}_4(i).
\end{equation}
The numerical values  $\mathcal{C}_\triangle$ and $\mathcal{C}_\square$ reported in Table \ref{Table1} were calculated using the networkx (2.6.3) package \cite{NetworkX,SciPyProceedings_11}.  In addition, we include the global times $\tau(0)$ in Eq. (\ref{globaltime_tau})  for the random walk dynamics on the networks without damage, this value gives an estimate of the number of steps necessary to reach any node from any initial condition. We also include the values $\langle\mathcal{F}(T^\star)\rangle/\rho_{\mathcal{E}}$ depicted in Fig. \ref{Fig_4}(b).
\\[2mm]
The values presented in Table \ref{Table1} are sorted with the values $\langle\mathcal{F}(T^\star)\rangle/\rho_{\mathcal{E}}$ and reveal different aspects of the networks and their response under cumulative damage measured by this normalized functionality. First, we see that in many cases, a higher number of $N$ or $|\mathcal{E}|$ does not imply a better response to damage.  In this respect, the fraction of triangles and squares in each network is given by $\mathcal{C}_\triangle$, $\mathcal{C}_\square$. Using these values we can identify all the networks with a tree-like structure (in the networks analyzed $\mathcal{C}_\triangle=0$, $\mathcal{C}_\square=0$), networks with only one triangle (systems 5, 7, 8) or only one square (systems 9, 10, 13). The systems with the strongest tolerance to damage  $\langle\mathcal{F}(T^\star)\rangle/\rho_{\mathcal{E}}>5$ (systems 21-33) include a major proportion of squares and triangles. In particular, the two most robust systems are the metro networks in Paris and London with similar values of $\langle\mathcal{F}(T^\star)\rangle/\rho_{\mathcal{E}}$. Here it is worth noticing that, in comparison with the metro in London, the metro in Paris has a lower number of nodes but more edges and a higher proportion of triangles.
In general, the proportion of triangles and squares is an important feature that may increase the tolerance to damage. However, it is important to notice that the value $\langle\mathcal{F}(T^\star)\rangle/\rho_{\mathcal{E}}$ also includes information of cycles with other sizes. For example, we have Hong Kong with one triangle and a cycle with five nodes.
\\[2mm]
\begin{figure*}[t!]
	\begin{center}
		\includegraphics*[width=1.0\textwidth]{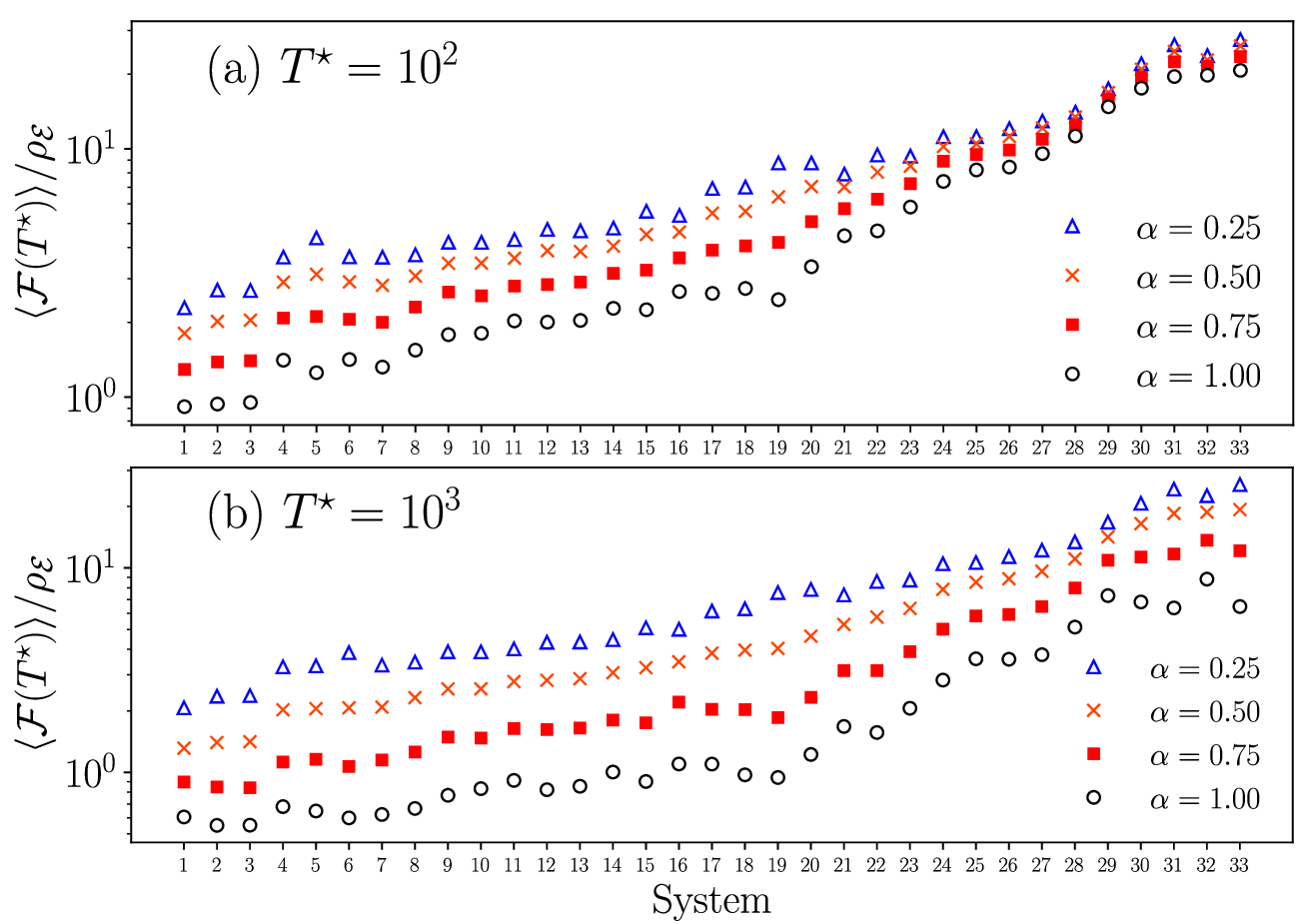}
	\end{center}
	\vspace{-5mm}
	\caption{\label{Fig_5} $\langle\mathcal{F}(T^\star)\rangle/\rho_{\mathcal{E}}$ for metro systems as sorted in Fig. \ref{Fig_4} for different values of $T^\star$ and $\alpha=0.25,\,0.50,\,0.75,\,1.00$. The results are obtained with 1000 realizations using Monte Carlo simulations of the algorithm of damage accumulation for (a) $T^\star=10^2$ and (b) $T^\star=10^3$. }
\end{figure*}

Finally, the time $\tau(0)$ provides additional information about the network topology characterizing the transport in the network without damage considering all the possible paths that connect two nodes through mean first passage times [see Eq. (\ref{TauGlobalMFPTdef})]. Although this global time helps us to define $\mathcal{F}(T)$, it does not characterize the network's resistance to damage. We can see this in Paris and London for which $ \tau(0)$ times differ significantly but have similar $\langle\mathcal{F}(T^\star)\rangle/\rho_{\mathcal{E}}$.
\\[2mm]
Let us now explore the ranking of the response to damage of metro systems for different parameters implemented in the Monte Carlo simulations. To this end, in Fig. \ref{Fig_5} we evaluate $\langle\mathcal{F}(T^\star)\rangle/\rho_{\mathcal{E}}$ for different values of $\alpha$ and $T^\star$, ensemble averages are calculated using $1000$ realizations. We then calculate the Kendall’s tau rank correlation $\mathcal{K}_{\mathrm{corr}}$. This rank correlation coefficient evaluates the degree of similarity between two sets of ranks given to a same set of
objects. The value $\mathcal{K}_{\mathrm{corr}}$ depends upon the number of inversions
of pairs of objects which would be needed to transform one rank
order into the other \cite{kendall1990correlation}. In this manner, we can compare the classifications of the metro systems (ranking) sorting them in increasing order of the respective values of $\langle\mathcal{F}(T^\star)\rangle/\rho_{\mathcal{E}}$ presented in Fig. \ref{Fig_5} for $T^\star=100,\, 1000$ and $\alpha=0.25,\, 0.5,\, 0.75,\, 1.0$. The results were obtained using as reference the ranking in Fig. \ref{Fig_4}(b) and also presented in Table \ref{Table1}, the values for the correlations with the ranking generated with $T^\star=100$ are $\mathcal{K}_{\mathrm{corr}}=0.95,\, 0.98,\, 0.98,\,0.97$ (for $\alpha=0.25,\, 0.5,\, 0.75,\, 1.0$); similarly, for the ranking with 
$T^\star=1000$, $\mathcal{K}_{\mathrm{corr}}=0.97,\, 1.0,\, 0.94,\, 0.89$. These results indicate that the rankings produced sorting the values of $\langle\mathcal{F}(T^\star)\rangle/\rho_{\mathcal{E}}$  are in closer agreement.
\\[2mm]
All the results in Table \ref{Table1} and Fig. \ref{Fig_5} show that the normalized value $\langle\mathcal{F}(T^\star)\rangle/\rho_{\mathcal{E}}$ is a good measure that characterizes the complexity of the structure and the global response of a network to accumulated damage and allows us to classify networks whose sole purpose is to communicate to all nodes. The application explored in this paper occurs in the context of the analysis of infrastructure in urban transportation systems. However, a similar framework can be implemented to analyze the vulnerability of different systems such as energy transport infrastructure, information networks, the transport of nutrients and oxygen in tissues, among many others.
\section{Conclusions}
In this research, we implement a model to evaluate the reduction of transport on a network due to the accumulation of damage in each of its edges. We analyzed 33 metro systems worldwide in which each link represents a group of stations in the same line of the metro. The nodes represent stations where users can change between lines or the end stations of a line. The studied networks present varied topologies that range from tree structures to networks with a higher fraction of cycles with three and four nodes.
\\[2mm]
We evaluate the evolution of transport through the ensemble average $\langle\mathcal{F}(T)\rangle$, its value decreases with $T$ that quantifies the total number of damage hits in a structure. We see that $\langle\mathcal{F}(T)\rangle$ describes the evolution of cumulative damage and is unique to each metro system. By using this information, we compare the metro systems considering the normalized functionality $\langle\mathcal{F}(T^\star)\rangle/\rho_{\mathcal{E}}$ in a particular time $T^\star$, this value considers the average effect of damage per edge and allows characterizing
the network topology along with its response under cumulative damage. Our findings using Monte Carlo simulations show how the robustness of the networks increases when
multiple paths can connect two nodes. We have demonstrated in the present paper that the robustness of a transportation network is a complex interplay of its topology (redundant paths between nodes) and its capacity to endure damage in a link (due to maintenance and reparation) which we described with the parameter $\alpha$.
\\[2mm]
The methods explored in this research are general and can be implemented to analyze different systems whose exclusive function is transport. These systems can describe other transportation modes in cities, information networks, energy transport, transport of nutrients in a tissue or extended to the analysis of other dynamical processes in complex systems, for example, synchronization. 
\\[2mm] It would be interesting to seek a general principle able to predict the effect of damage directly from the structure of the network without implementing Monte Carlo simulations; a generalization of this type would require combining techniques of random matrix theory and spectral graph theory. 

\section*{Acknowledgments}
LKEH acknowledges support from CONACYT M\'exico. LKEH and APR acknowledge support from Ciencia de Frontera 2019 (CONACYT), project ``Sistemas complejos estoc\'asticos: Agentes m\'oviles, difusi\'on de part\'iculas, y din\'amica de espines'' (Grant No. 10872).

\section*{References}

\providecommand{\noopsort}[1]{}\providecommand{\singleletter}[1]{#1}%

\end{document}